# Core Challenge 2023
# Solver and Graph Descriptions

Edited by

Takehide Soh
Kobe University, Japan

Tomoya Tanjo
National Institute of Genetics, Japan

Yoshio Okamoto
The University of Electro-Communications, Japan

Takehiro Ito
Tohoku University, Japan

# Contents

# *recongo-isrp*: an ASP-based Independent Set Reconfiguration Solver

Masato Kato, Shuji Kosuge, Yuya Yamada, Kazuki Takada, Aoi Ito, and Mutsunori Banbara

Graduate School of Informatics, Nagoya University, Japan

*Recongo-isrp* is a system for solving the Independent Set Reconfiguration Problem (ISRP; [2]) based on Answer Set Programming(ASP; [1]). The *recongo-isrp* solver reads an ISRP instance of DIMACS format and converts it into ASP facts. In turn, these facts are combined with an ASP encoding for ISRP solving, which can subsequently be solved by an ASP-based combinatorial reconfiguration solver *recongo*. The *recongo* solver is implemented by using python interface of an efficient ASP solver *clingo* [3]. The high-level approach of ASP has obvious advantages. Main features of *recongo-isrp* are as follows:

- ASP provides a rich language and is well suited for modeling combinatorial reconfiguration problems.
- The *recongo-isrp* encoding for ISRP solving is compact and human-readable.
- The *recongo-isrp* encoding can be extensible for capturing new additional constraints.
- ISRPs are solved by a general-purpose solver rather than dedicated implementation.

For Core Challenge 2023, we implement new version of *recongo* solver by utilizing the latest python interface of *clingo*. We also develop an ASP encoding for checking the unreachability of ISRP based on *numeric abstraction* proposed by Remo Christen et al (PARIS team of Core Challenge 2022).

# PARIS 2023: Planning Algorithms for Reconfiguring Independent Sets

Remo Christen[1], Salomé Eriksson[1], Michael Katz[2], Christian Muise[3], Florian Pommerening[1], Jendrik Seipp[4], Silvan Sievers[1], and David Speck[4]

[1]University of Basel
[2]IBM T.J. Watson Research Center
[3]Queen's University
[4]Linköping University

## 1 Introduction

In this report, we briefly describe our entry to the 2023 ISR competition: Planning Algorithms for Reconfiguring Independent Sets (PARIS 2023). Our solver is a modified version of the 2022 competition submission, which performed extremely well across several of the tracks Soh et al. [2022]. We have adapted the solver given the newly imposed resource limits and implemented a mechanism for the portfolio approach to return the best solution found during the resource limits. We additionally employ a suite of anytime search methods, which may produce better solutions. Careful handling of the time-limits was required to ensure that the solver responds with an answer in time. In the following, we describe the components of our planner and how we combine them for the different tracks.

## 2 Components

Throughout the portfolio tracks, we use three core planners: (1) a planner specializing in finding short plans; (2) a planning specializing in finding long plans; and (3) a solver dedicated to proving unsolvability of the ISR instances. In the single solver tracks, we pick one of the planners for each track. In this section, we first describe each component separately, then discuss how they are combined in the next section. The first two components are classical planners and we transform the given col/dat files to planning task inputs in the same way we did in 2022.

## 2.1 Using Landmarks to Find Short Plans

Similar to the 2022 entry, we run a greedy best-first search (GBFS) Doran and Michie [1966] with a *landmark count* heuristic Keyder et al. [2010] that computes all landmarks of the *delete-relaxed* task Bonet and Geffner [2001] ($h^1$ landmarks). The landmark costs are combined with *uniform cost partitioning* Katz and Domshlak [2008]. In addition to this configuration that also ran in 2022, we consider an anytime version that does not stop after finding the first solution but instead continues to search for shorter and shorter plans. The configuration runs weighted $A^*$ with decreasing weights, starting with GBFS (which is equivalent to using an infinitely large weight), then using weights of 5, 3, 2 and 1 (the last of which is equivalent to running $A^*$). This anytime configuration will continually find shorter and shorter plans as long as the time allows.

## 2.2 Using Symbolic Top-$k$ Search to Find Long Plans

As in 2022, we run a modified forward symbolic blind search Torralba et al. [2017], Speck et al. [2020a] based on an algorithm called SymK-LL von Tschammer et al. [2022], implemented in the symbolic planner SymK Speck et al. [2020b], which iteratively finds and generates all loopless plans of a given task. This configuration of SymK finds increasingly longer loopless plans, starting with the shortest plan. It can be seen as an anytime configuration like our landmark component, but it starts from the shortest plan and approaches the longest loopless plan whereas the landmark configuration starts with a suboptimal plan and approaches the shortest plan. In addition to this anytime configuration, we also use an optimal configuration that stops after the first solution is found.

## 2.3 Detecting Unsolvability with Counter Abstractions

Again, as in 2022, we use abstractions based on counting how many tokens are on nodes with a particular color (in some coloring of of the graph). An abstract state is given by a count specifying for each color how many tokens are on nodes with that color. We use a MIP solver to test whether an independent set is possible for a given abstract state. If it is not, we can prune the abstract state. Using this pruning, we fully explore the abstract state space of the counter-abstraction. If no abstract goal state is reachable, we have a proof that the instance is unsolvable. We refer to the 2022 for a longer description of this component. Since the component uses CPLEX Cplex [2009] and we cannot publish it for licensing reasons, we implemented a fall back solution using SCIP Achterberg [2009] for it. Our container can be built with CPLEX support if a CPLEX installer is available.

# 3 Configurations

For all our configurations, we reserve 60 seconds at the end to report the best solution found and 200 MB of memory for the Python driver script starting the solvers. This is necessary to orchestrate the multiple simultaneous threads being used to find a solution. Reserving time to pick the best solution is necessary even for single solver tracks, as we run anytime solvers that continually find better and better solutions. In portfolio configurations, we divide the available memory equally among all components and let them all run for the full time. If a component finishes, we decide based on its result if it is necessary to keep the other components running. Our two planner components (SymK and the one based on landmarks) are single-core applications, while the counter abstraction component uses a MIP solver that can use multiple cores. We do not explicitly enforce this but if we start all of them in parallel, the operating system will schedule the two planners to run on one core each and the MIP solver to use the remaining available cores.

## 3.1 Shortest Track

For the single solver configuration, we only run the anytime version of the landmark-based planner. It finds successively better plans, and given sufficient resources, it will find the optimal plan. For the portfolio configuration, we additionally run the optimal configuration of SymK. We do not run the counter abstraction component, since no points are granted for identifying unsolvable instances. If SymK terminates in this configuration, it means the plan it found is optimal. In this case, we terminate the other component. If the anytime search runs to completion, we also know that the last plan it found was optimal and we can terminate SymK. If none of these cases occur, we run until the time limit and report the best solution found up until that point.

## 3.2 Existent Track

Our configuration for the existent track is almost the same as in the shortest track. The difference is that we include the counter abstraction and terminate all components as soon as any component finds the first solution (or we prove unsolvability). This means we replace the anytime configuration with one that only runs until it finds the first solution. The single solver configuration just runs this configuration of the landmark-based planner, and the portfolio configuration also runs the optimal configuration of SymK, and the MIP solver.

## 3.3 Longest Track

In the single solver configuration, we run the anytime configuration of SymK that keeps finding longer and longer plans. Once we reach the resource limits, we report the longest solution found. In the portfolio configuration, we also run the landmark-based planner in parallel. We do not use its anytime configuration here because it would find shorter and shorter plans, so the first plan it finds is the longest one it can find. The landmark planner is generally optimized for finding short solutions, but in case SymK does not find any solution, or does not find long solutions fast enough, a solution by this component could still be better.

# RECONF-BFS

Masaya Sano[1]

[1]Shinshu University

## 1 Description

The solver is based on breadth-first search.

# A decision diagram-based solver for the independent set reconfiguration problem

Jun Kawahara[1] and Shou Ohba[1]

[1]Kyoto University

## 1 Engine: Core Solver or Algorithm

Our solver is based on the algorithm [2] that uses zero-suppressed binary decision diagrams [4] and the graphillion library [1].

## 2 ZDD

A ZDD is a data structure for efficiently representing a family of sets. The definition of a ZDD is given below. Let $X = \{x_1, \ldots, x_n\}$ be an underlying set of a family represented by a ZDD, and let $x_1 < x_2 < \cdots < x_n$. A ZDD is a directed acyclic graph (DAG) $D = (N, A)$ with the following property. $D$ has at most two nodes with outdegree zero, which are called 0-terminal and 1-terminal. Nodes other than the terminals are called non-terminal nodes. A non-terminal node has one element of $X$ and is called a label. A non-terminal node has two arcs, called a 0-arc and a 1-arc. If a non-terminal node $\nu \in N$ has label $x \in X$ and its 0-arc and 1-arc point at $\nu_0, \nu_1 \in N$, respectively, we write $\nu = (x, \nu_0, \nu_1)$. If $\nu_i = (x^i, \nu_{i0}, \nu_{i1})$ holds, $x < x^i$ must be satisfied $(i = 0, 1)$. $D$ has exactly one node with indegree zero, called the root.

A node $\nu$ in a ZDD recursively represents a family of sets as follows. If $\nu$ is the 0- and 1-terminal, it represents $\emptyset$ and $\{\emptyset\}$, respectively. If $\nu$ is a non-terminal node, we write $\nu = (x, \nu_0, \nu_1)$. The family $\mathcal{S}(\nu)$ of sets is the union of the family of sets represented by $\nu_0$ and the family of sets obtained by adding $x$ to each element of the family of sets represented by $\nu_1$. That is, $\mathcal{S}(\nu) = \mathcal{S}(\nu_0) \cup (\{\{x\}\} \bowtie \mathcal{S}(\nu_1))$, where $\mathcal{F} \bowtie \mathcal{G} = \{F \cup G \mid F \in \mathcal{F}, G \in \mathcal{G}\}$. We interpret the family of sets represented by its ZDD as the family of sets represented by the root node. For more information on ZDDs, please refer to [3].

# 3 Algorithm for solving the independent set reconfiguration problem

Given a graph $G = (V, E)$, an independent set of $G$ can be represented as a subset of $V$. The collection of all independent sets of $G$ can be represented by a family of sets with underlying set $V$. For a family $\mathcal{F}$ of sets with underlying set $V$ and a set $A \in V$, let $\mathsf{swap}(\mathcal{F}, A)$ be the family of all the sets obtained by removing an element from every set in $\mathcal{F}$ and adding an element in $A$ to the set. That is,

$$\mathsf{swap}(\mathcal{F}, A) = \{F \cup \{x\} \setminus \{x'\} \mid F \in \mathcal{F}, x \notin F, x \in A, x' \in F\}.$$

Our algorithm for computing $\mathsf{swap}(\mathcal{F}, A)$ is described in the next section. Using $\mathsf{swap}$, the token-jumping model of the independent set reconfiguration problem can be solved as follows. In what follows, the underlying set of all families is $V$. Let $\mathcal{F}_{\text{ind}}$ be the family of all the independent sets of $G$. Let $\mathcal{F}_0 = \{S\}$. We recursively define $\mathcal{F}_{i+1} = \mathsf{swap}(\mathcal{F}_i, V) \cap \mathcal{F}_{\text{ind}}$. $\mathsf{swap}(\mathcal{F}_i, V)$ is the family of sets obtained by adding and removing one element of each set of $\mathcal{F}_i$. By taking the intersection of $\mathsf{swap}(\mathcal{F}_i, V)$ and $\mathcal{F}_{\text{ind}}$, we extract independent sets from $\mathsf{swap}(\mathcal{F}_i, V)$. $\mathcal{F}_i$ is the family of independent sets obtained by reconfiguring $S$ using at most $i$ steps (satisfying that the intermediate sets are also independent sets). The above computation is performed for $i = 0, 1, \ldots$ in this order. If $T \in \mathcal{F}_i$ holds for some $i$, we know that there exists a reconfiguration sequence from $S$ to $T$ with length $i$, and the algorithm stops. If $\mathcal{F}_i = \mathcal{F}_{i+1}$ for some $i$, no new independent set is obtained, and we know that there is no reconfiguration sequence from $S$ to $T$. From the construction of $\mathcal{F}_i$, the smallest $i$ such that $T \in \mathcal{F}_i$ holds is the minimum number of steps among all the reconfiguration sequences from $S$ to $T$.

# 4 Reconfiguration operation for families of sets using ZDDs

In this section, we describe how to realize the algorithm described in the previous section using ZDDs. A method to construct a ZDD representing $\mathcal{F}_{\text{ind}}$ is described in the book [3]. The construction of a ZDD for $\mathcal{F}_0 = \{S\}$ is obvious. For any set $A$, there exists a method to determine whether $A$ belongs to $\mathcal{F}$. Given two families $\mathcal{F}, \mathcal{F}'$ of sets as ZDDs, a method for constructing a ZDD representing $\mathcal{F} \cap \mathcal{F}'$ has been proposed [4]. Therefore, if there is a method for constructing a ZDD representing $\mathsf{swap}(\mathcal{F}, A)$, all the above procedures can be performed as ZDD operations.

Given a family $\mathcal{F}$ of sets as a ZDD, we describe how to construct a ZDD representing $\mathsf{swap}(\mathcal{F}, A)$. The $\mathsf{swap}$ operation uses the following operations:

$$\mathsf{rem}(\mathcal{F}) = \{F \setminus \{x\} \mid F \in \mathcal{F}, x \in F\},$$
$$\mathsf{add}(\mathcal{F}, A) = \{F \cup \{x\} \mid F \in \mathcal{F}, x \notin F, x \in A\}.$$

For every $x \in X$, we can write $\mathcal{F} = \mathcal{F}_0 \cup (\{\{x\}\} \bowtie \mathcal{F}_1)$. We let $\mathcal{F}^{\text{rem}} = \mathsf{rem}(\mathcal{F})$ and $\mathcal{F}^{\text{rem}} = \mathcal{F}_0^{\text{rem}} \cup (\{\{x\}\} \bowtie \mathcal{F}_1^{\text{rem}})$, where any set in $\mathcal{F}_0^{\text{rem}}$ and $\mathcal{F}_1^{\text{rem}}$ does not contain $x$.

Each element of $\{\{x\}\} \bowtie \mathcal{F}_1^{\text{rem}}$ is obtained by removing one element other than $x$ from each set in $\{\{x\}\} \bowtie \mathcal{F}_1$. Thus, $\mathcal{F}_1^{\text{rem}} = \mathsf{rem}(\mathcal{F}_1)$ holds. Also, each element of $\mathcal{F}_0^{\text{rem}}$ is obtained by removing one element from each set in $\mathcal{F}_0$, or by removing $x$ from each set in $\{\{x\}\} \bowtie \mathcal{F}_1$. Thus, $\mathcal{F}_0^{\text{rem}} = \mathsf{rem}(\mathcal{F}_0) \cup \mathcal{F}_1$ holds.

We describe our algorithm $\mathsf{rem}(\nu)$ that constructs a ZDD representing the family $\mathsf{rem}(\mathcal{S}(\nu))$ of sets for a ZDD $\nu = (x, \nu_0, \nu_1)$. If $\nu$ is the 0- or 1-terminal, the 0-terminal is returned. Otherwise, we perform the following. First, recursively compute $\mathsf{rem}(\nu_0)$ and take the union of it and $\nu_1$ (as a ZDD operation) [3]. Next, recursively compute $\mathsf{rem}(\nu_1)$. Finally, return $(x, \mathsf{rem}(\nu_0) \cup \nu_1, \mathsf{rem}(\nu_1))$ as the root node of $\mathsf{rem}(\nu)$.

For a ZDD $\nu = (x, \nu_0, \nu_1)$, the algorithm $\mathsf{add}(\nu, A)$, which constructs a ZDD representing $\mathsf{add}(\mathcal{S}(\nu), A)$, can be written as well as $\mathsf{rem}$ in a recursive mannar. If $\nu$ is the 0-terminal, it returns the 0-terminal. If $\nu$ is the 1-terminal, it returns the ZDD representing $\{\{x\} \mid x \in A\}$ (the construction of the ZDD is clear). If $x < \min(A)$, return $(x, \mathsf{add}(\nu_0, A), \mathsf{add}(\nu_1, A))$ because $x < \min(A)$ means $x \notin A$, where $\min(A)$ is the element $y$ in $A$ such that for all $y'(\neq y) \in A$, $y < y'$ holds. If $x = \min(A)$, return $(x, \nu_0', \nu_1')$, where $\nu_0' = \mathsf{add}(\nu_0, A \setminus \{x\})$, and $\nu_1' = \mathsf{add}(\nu_1, A \setminus \{x\}) \cup \nu_0$. If $x > \min(A)$, return $(\min(A), \mathsf{add}(\nu, A \setminus \{x\}), \nu)$.

For a ZDD $\nu = (x, \nu_0, \nu_1)$, the algorithm $\mathsf{swap}(\nu, A)$, which constructs a ZDD representing $\mathsf{swap}(\mathcal{S}(\nu), A)$, can be written as well as $\mathsf{rem}$ and $\mathsf{add}$ in a recursive mannar. If $\nu$ is the 0- or 1-terminal, it returns the 0-terminal. If $x < \min(A)$, return $(x, \mathsf{swap}(\nu_0, A), \mathsf{swap}(\nu_1, A))$. If $x = \min(A)$, return $(x, \nu_0', \nu_1')$, where $\nu_0' = \mathsf{swap}(\nu_0, A \setminus \{x\}) \cup \mathsf{add}(\nu_1, A \setminus \{x\})$, and $\nu_1' = \mathsf{swap}(\nu_1, A \setminus \{x\}) \cup \mathsf{rem}(\nu_0)$. If $x > \min(A)$, return $(\min(A), \mathsf{swap}(\nu, A \setminus \{x\}), \mathsf{rem}(\nu))$.

# 5 Update from CoRe Challenge 2022

The difference from the solver used in CoRe Challenge 2022 is the use of *restrict* operation. For two ZDDs $\mathcal{F}$ and $\mathcal{G}$ and an integer $k$, the restrict operation is defined as

$$\mathsf{restrict}(\mathcal{F}, \mathcal{G}, k) = \{F \in \mathcal{F} \mid \exists G \in \mathcal{G}, G \subseteq F, |F - G| = k\}.$$

Using the restrict operation, we obtain

$$\mathcal{F}^{i+1} = \mathsf{restrict}(\mathcal{F}_{\text{ind}}, \mathsf{rem}(\mathcal{F}^i, 1), 1).$$

# ISR 2023 Graph Track Documentation

Bruce Chidley[1], Christian Muise[1], and Alice Petrov[1]

[1]Queen's University

## 1 Introduction

Several strategies were tried throughout the course of the contest, from hand-crafted examples to exhaustive enumeration. Ultimately, small graphs with a predicable structure, referred to as "widgets", provide a useful pattern for repeated movements. Since connecting these graphs is readily scalable, this is what is used for the submission. Throughout the remainder of the document, we describe the construction of these widgets and the final graphs.

### 1.1 5 Node Widget

We leverage a five node subgraph we call the "house widget": a 4-cycle with two adjacent nodes leading to a 5th (essentially, a triangle sitting on top of a square). The house widget has a number of properties that make it ideal to use as a building block in creating exponential sequences.

1. The graph has an optimal "long" shortest reconfiguration sequence for ISR instances of order 5.
2. You can only place two tokens on this widget, meaning each step of the reconfiguration sequence consists of a maximum independent set. In other words, the sequence is "tight" and no additional nodes can be added to the independent set at any point.
3. The topmost node, which we call the "anchor", is occupied throughout the entire sequence with the exception of the starting state and ending state, and is required to switch the corners that the two tokens are on.
4. The sequence is unique. Thus, the solution space is a path and the behaviour of the widget is predictable.

We call the start state "on" and the symmetrically opposite state "off". Note that the anchor is both required to switch a house from "on" to "off" and is occupied throughout the sequence.

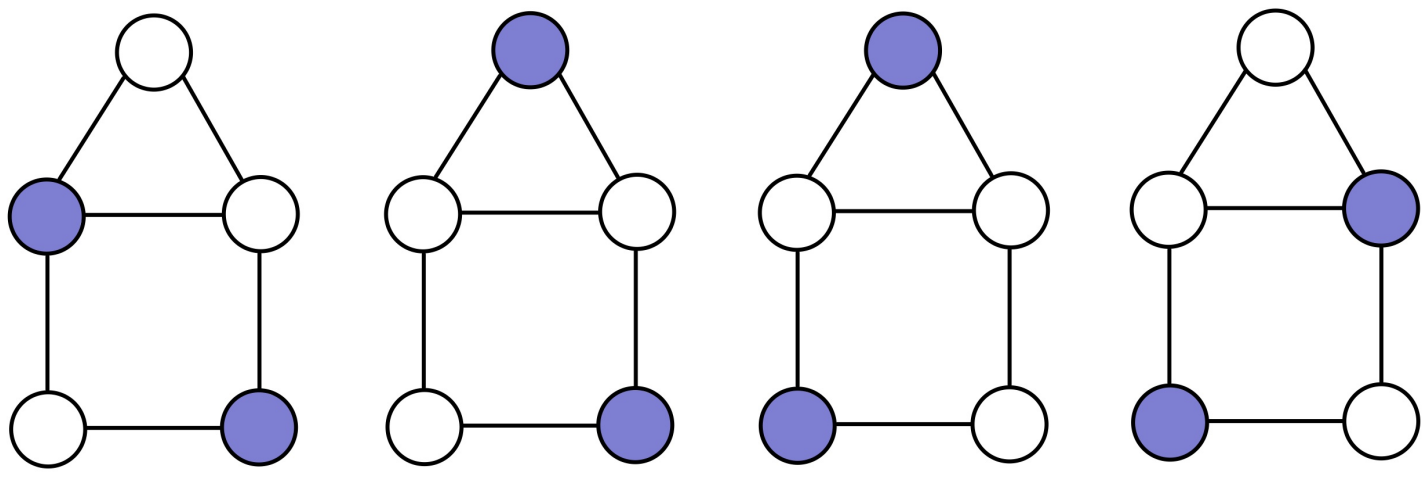

Figure 1: Reconfiguration sequence from “off” to “on”

## 1.2 10 Node Widget

The 10 node widget, introduced by the @tpierron team at the CoRe 2022 challenge, is an optimal solution for the reconfiguration problem in the n = 10 case. The graph was obtained by brute forcing all graphs on n = 10 vertices.

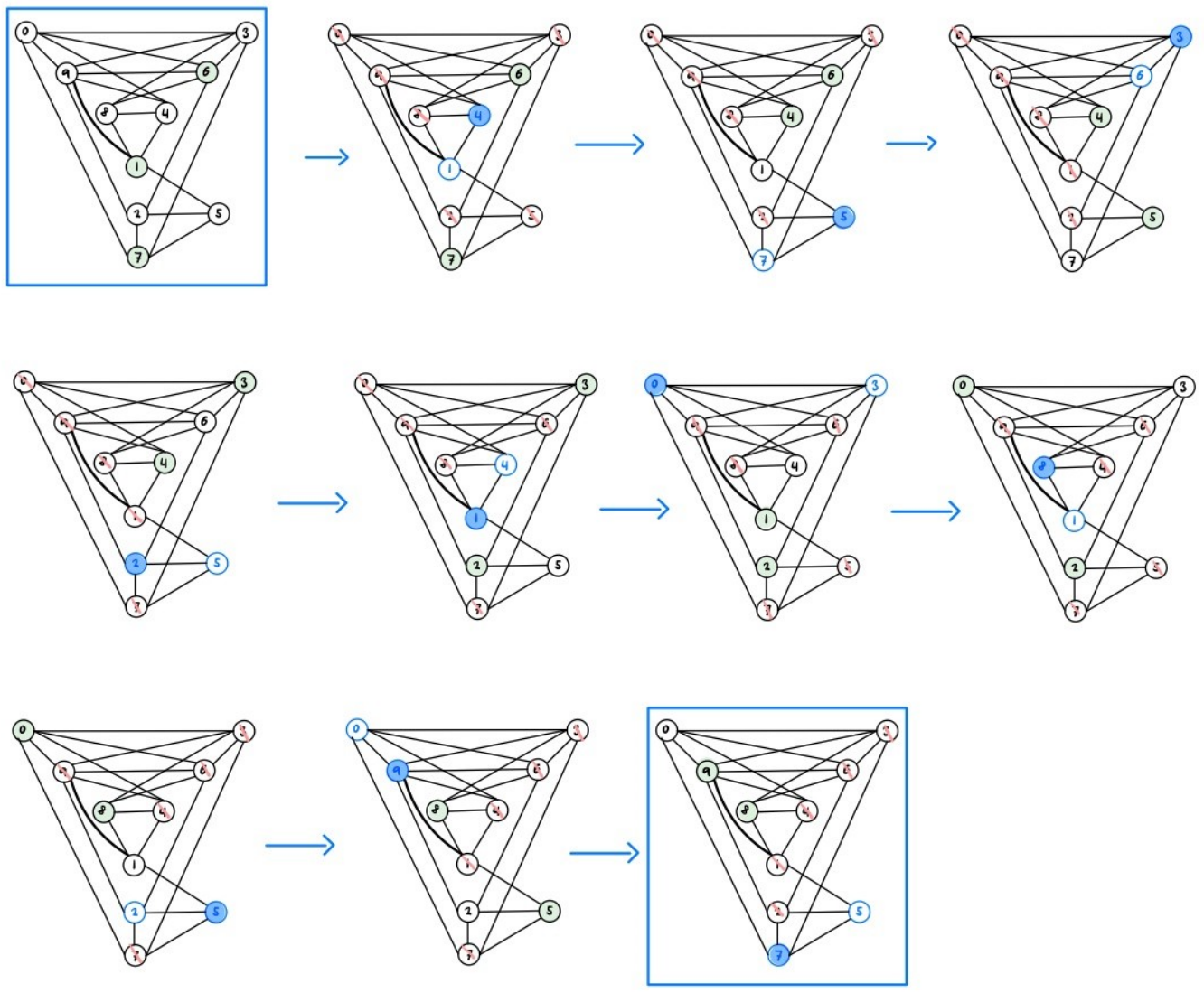

Figure 2: Reconfiguration sequence of the 10 node widget

# 2 Submissions

## 2.1 17 Node Graph

This graph uses a 10 node widget, a house widget, and one “2 node widget”, consisting of two vertices (A and B) connected by a single edge. Note that in the reconfiguration sequence of the 10 node widget, we place tokens on vertices 1 and 5 two times each, for a total of four times. The graph is constructed as follows: In order to use vertex 1, the house must be “on”. In order to use vertex 5, the house must be “off”. To enforce this, we connect vertex 1 to the two vertices in the “off” state and vertex 5 to the two vertices in the “on”state. Vertex A of the “2 node widget” is connected to the anchor, and vertex B is connected to vertices 1 and 5 in the 10 node widget. Thus, in the resulting graph, in order to flip the 10 node widget, we must flip the house and 2 node widget every time we hit nodes 1 or 5. To see why the entire sequence is enforced, note that node A is connected to the anchor, and node B is connected to both 1 and 5, which means that in order to use node 1 or 5, the 2-node widget needs to be on node A and thus the anchor needs to be free. This results in 10 steps to flip the 10 node widget, plus 4 flips of length 5 to flip the house and 2 node widgets, for a total reconfiguration sequence of length 30.

## 2.2 31 Node Graph

This graph is comprised of three 10 node widgets and one left over node. The three 10 node widgets were first joined together per the algorithm outlined by the @tpierron team, resulting in a 30 node graph with an expected corresponding path length of 210. Then, the one leftover node was added into the graph by putting it to be adjacent to all other nodes except the end state of the just described 30 node graph. This way, when the 30 node graph’s path is complete, one of the occupied nodes comprising the end state can move to the final leftover node. This method results in a total reconfiguration sequence of length 211.

## 2.3 59 Node Graph

This graph is made up of five 10 node widgets, one house widget, and one “4 node widget”, consisting of four vertices (A, B, C, and D) connected in a chain from A to D. First, the five 10 node widgets were joined together as per the algorithm outlined by the @tpierron team. This results in a 50 node graph whose “long” shortest path length is 3410. Then, the house widget was added in, serving as a doubling widget. This was done by first connecting the anchor of the house to all nodes in the just created 50 node graph except for its end state. This makes it so that the house can only flip bits when the 50 node graph is in its end state. Then, by setting the overall start state to be the start state of the 50 node graph plus the “off” position of the house, and having the overall end state be the start state of the 50 node graph plus the ”on” position of the house, the 50 node graph’s path is traversed twice with three more steps due the house flip, effectively doubling its path length. The 50 node graph must go from its start position to its end position, at which point the house flips bits, and then the 50 node graph goes all the way back to its start position again, increasing the path length to 6823. Finally, the 4 node widget was added similarly to how the house widget and 2 node widget were added in the 17 node case. The 50 node graph is structured in such a way that there are five underlying 10 node graphs that all go from their respective start states to their end states some number of times, with the most recently added 10 node widget only flipping once, and the first added 10 node widget flipping the most. By exploiting this property, we can join nodes A and D to nodes 1 and 5 of the first added 10 node widget to have the 4 node widget flip every time those nodes appear in the path. Specifically, the 4 node widget will start with A and C occupied, and will always have either A and C, or B and D occupied. Then, A is put to be adjacent to node 5 of the first 10 node widget, and D is put to be adjacent to node node 1 of the first 10 node widget. Node D is also put to be adjacent to the anchor of the 5 node house widget so that one more flip is achieved when the house flips. This method results in a total reconfiguration sequence of length 9895.

# Instance Description for Core Challenge 2023

Akira Suzuki[1]

[1]Graduate School of Information Sciences, Tohoku University.

## 1 Description

Our construction is an improvement on the method used by the previous champion team "tpierron," for $n = 50$ and $n = 100$.

They constructed graphs with 50 or 100 vertices by combining multiple sets of a graph with 10 vertices each as a *piece*. However, in Core Challenge 2023, we needed to construct graphs with 17, 31, and 59 vertices, so we could not use their method as is.

Therefore, we created new pieces with 3 to 9 vertices. Then we used a dynamic programming algorithm to find the combination with the longest reconfiguration length among the graphs with 17, 31, and 59 vertices that can be constructed by combining these pieces. Figure 1 shows the pieces with 3 to 9 vertices.

Using a similar method, we generated graphs with 3 to 100 vertices. Figure 2 summarizes the reconfiguration lengths of the instances constructed using our method.

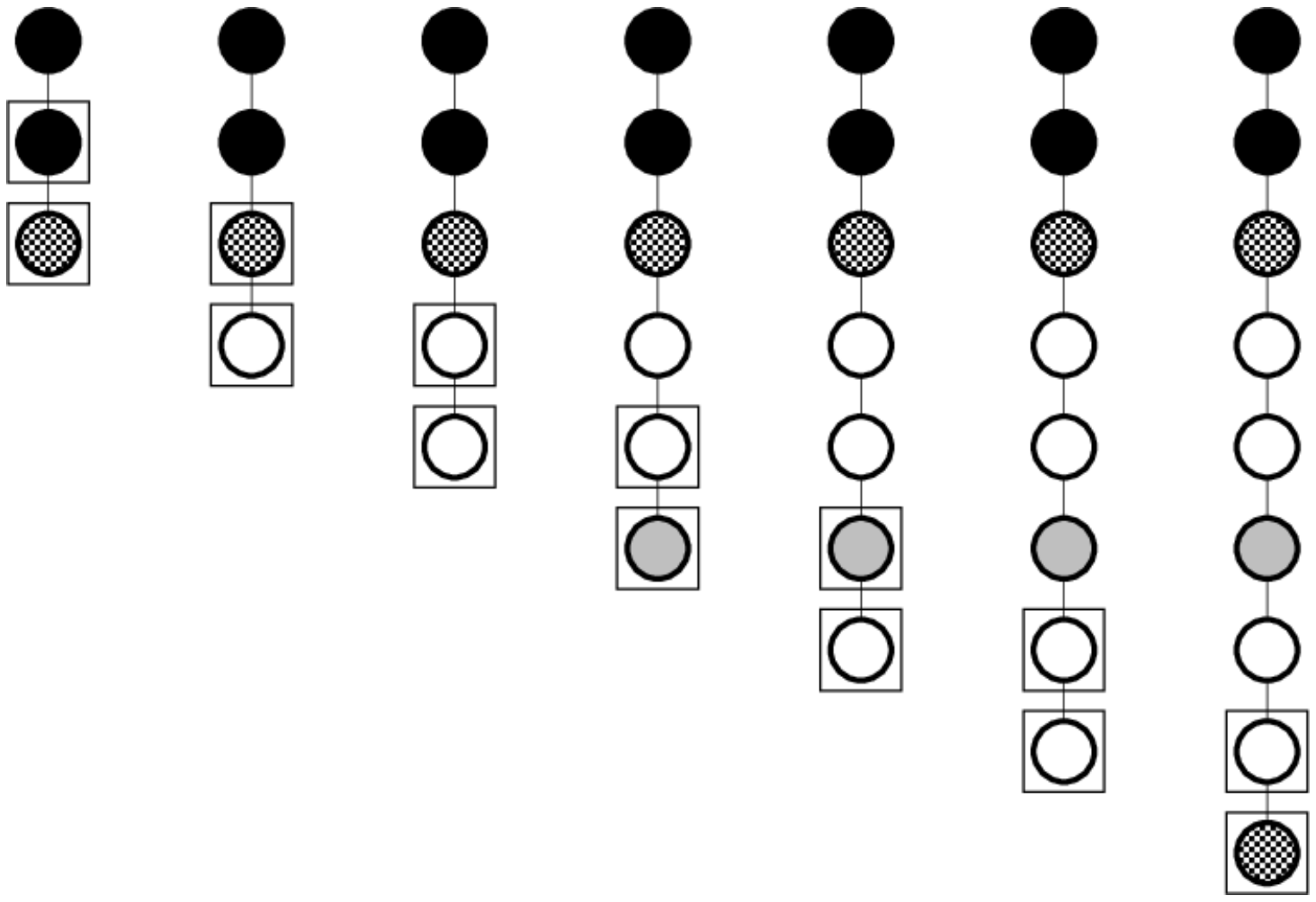

Figure 1: The pieces with 3 to 9 vertices. Note that these figures illustrates complement graphs, that is, where there is an edge, there is no edge, and where there is no edge, there is an edge. The black vertices are initial solution. The vertices enclosed by the square are the target solution. The gray vertices connect the initial solution to the original graph. The checked vertices connect to the target solution of the original graph.

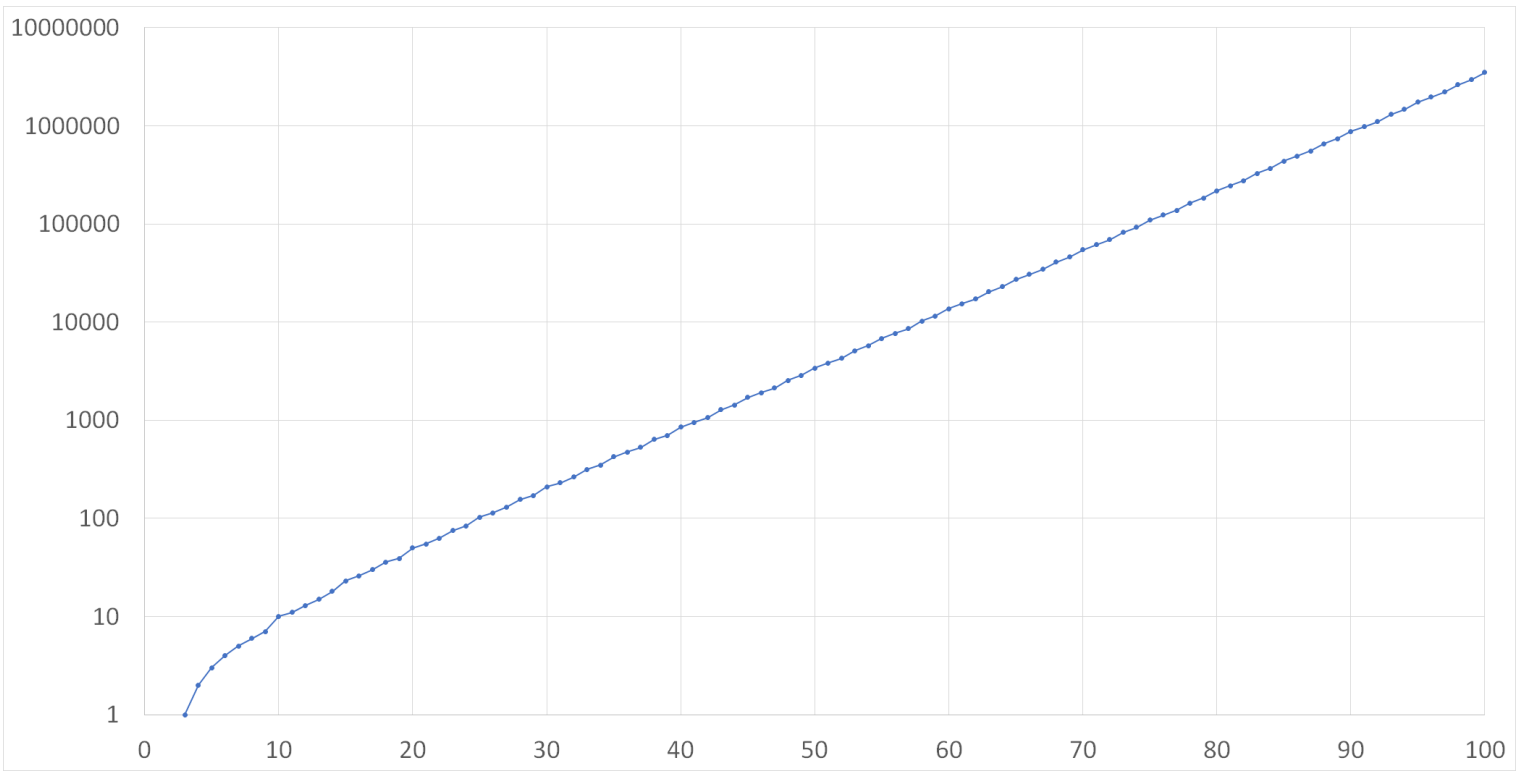

Figure 2: The reconfiguration lengths of the instances constructed using our method. The horizontal axis represents the number of vertices in the graph and the vertical axis represents the reconfiguration length.

# Synergy of Two Powerful Gadgets

Naoki Domon[1], Daichi Wakayama[1], and Takahiro Suzuki[1]

[1]Graduate School of Information Science, Tohoku University

## 1 Instance of (Graph#17) Track

- Our instance, as shown in figure 1, is handcrafted with respect to previous work [1]. The initial independent set is $\{1, 4, 6, 9, 12, 14, 17\}$(write $S_0$), and the target independent set is $\{1, 4, 6, 9, 12, 15, 17\}$(write $S_t$). It consits of 2 gadgets $X$,$Y$: $|X| = 12$, $|Y| = 5$. Gadget $X$ is based on the instance of (Graph#10) Track in [1], and we join two vertices 1,2 on the top of the gadget. $Y$ is well known gadget as "house gadget". Since maximal independent set size of $X$ is 5, it cannot happen that a token in $Y$ jumps to $X$. The same is true for $Y$, so we can consider gadget $X$ and $Y$ apart. We write $S_0 \cap X = S_{0X}$, $S_0 \cap Y = S_{0Y}$ (respectively $S_{tX}$, $S_{tY}$).
- The difference between initial set and target set is $\{14, 17\}$ and $\{15, 16\}$. First, we try to move a token on 14 to 13, but some of vertices in $X$ and adjacent to vertex 13 is filled with token, so we need to move these token to these non-adjacent to vertex 13. So, we define the target set of $X$ ($\{2, 3, 7, 9, 12\}$) $S_m$. It takes 13 steps from $S_{0X}$ to $S_m$.
- Now, we can jump a token to vertex 13, so we can reconfigurate tokens from $S_{0Y}$ to $S_{tY}$. After that, the last we need to do is reconfigurating $S_m$ to $S_{tX}$. It takes 13 steps, same as $S_{0X}$ to $S_m$. The total steps are $13 + 3 + 13 = 29$.

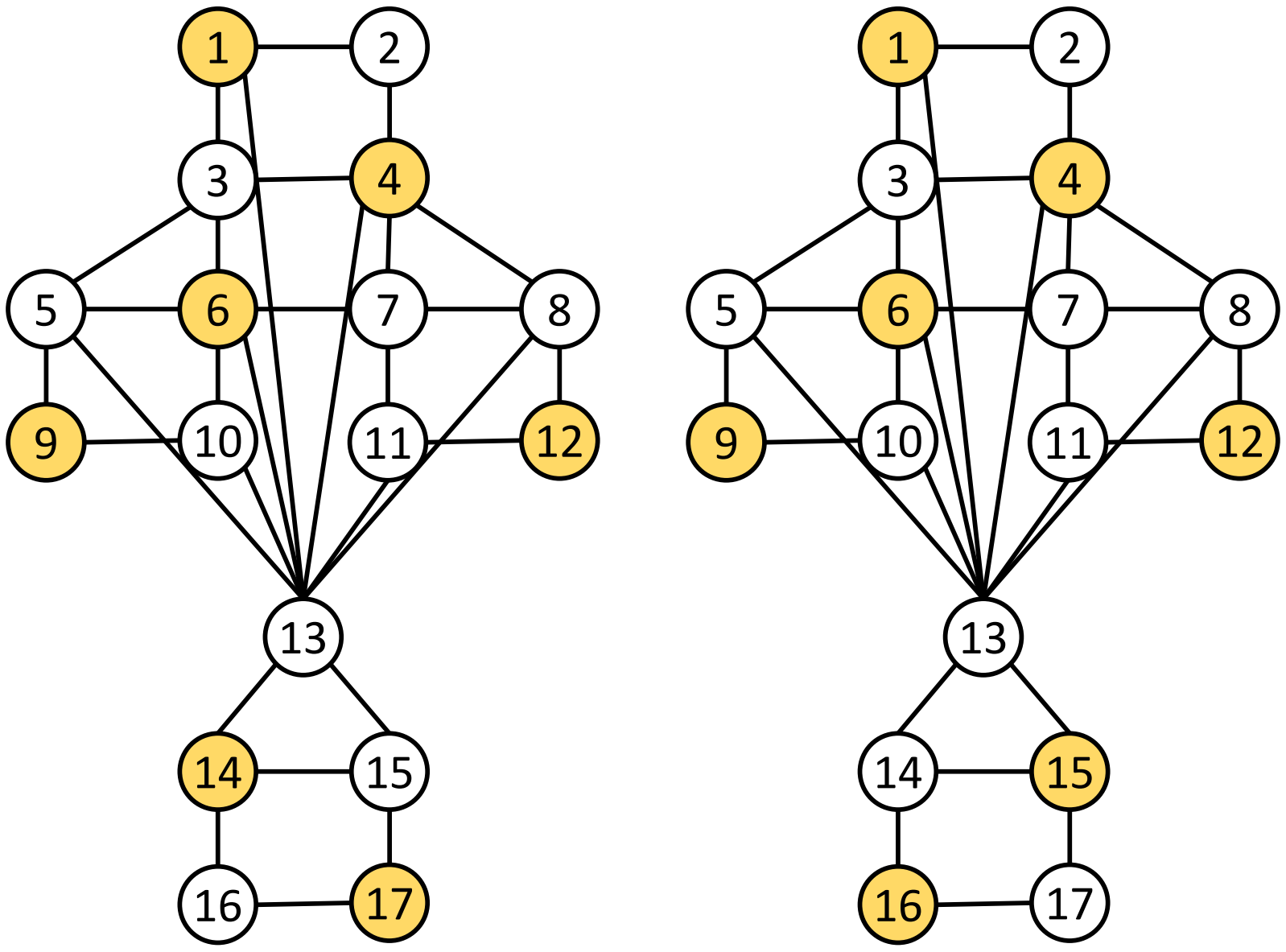

Figure 1: The instance of (Graph#17) Track, which takes 29 step. Initial independent set(left),and target independent set(right).